\documentstyle[aps,preprint,epsfig]{revtex}

\begin{document}
\title
{Measuring Double Parton Distributions in Nucleons at
Proton-Nucleus Colliders}

\author{M. Strikman}

\address{Department of Physics, Pennsylvania State University, University
Park, PA 16802}

\author{D.Treleani}

\address{ Universit{\`a} di Trieste; Dipartimento di Fisica Teorica,
Strada Costiera 11, Miramare-Grignano, and INFN, Sezione di
Trieste, I-34014 Trieste, Italy.}

\maketitle

\begin{abstract}
We predict a strong enhancement of multijet production in
proton-nucleus collisions at collider energies, as compared to a
naive expectation of a cross section $\propto A$. The study of the
process would allow to measure, for the first time, the double
parton distribution functions in a nucleon in a model independent
way and hence to study both the longitudinal and the transverse
correlations of partons.
\end{abstract}

\vspace{3cm}

E-mail strikman@color.phys.psu.edu \\

E-mail daniel@trieste.infn.it \\

\newpage
The systematic studies of hard inclusive processes during the last
two decades have let to a pretty good understanding of the single
parton densities in nucleons. However very little is known about
multiparton correlations in nucleons which can provide a new tool
for discriminating between different models of nucleons. Such
correlations may be generated, for example, by the fluctuations of
the transverse size of the color field in the nucleon leading, via
color screening, to correlated fluctuations of the densities of
gluons and quarks. A related source of correlations is QCD
evolution, since a selection of a parton with a given $x, Q^2$ may
lead to a local (in transverse plane) enhancement of the parton
density at different $x$ values.

It was recognized already more than two decades ago \cite{dpth}
that the increase of parton densities at small $x$ leads to a
strong increase of the probability of nucleon-nucleon collisions
where two or more partons of each projectile experience pair-vice
independent hard interactions. Although the production of
multijets through the double parton scattering mechanism was
investigated in several experiments \cite{dpex,cdf} at $pp,p\bar
p$ colliders, the interpretation of the data was hampered by the
need to model both the longitudinal and the transverse partonic
correlations at the same time. The aim of this letter is to point
out that the near future perspectives to study proton-nucleus
collisions at RHIC, as well as the plans for $pA$ collisions at
LHC, provide a feasible opportunity to study separately the
longitudinal and transverse partonic correlations in the nucleon
as well as to check the validity of the underlying picture of
multiple collisions.

The simplest case of a multiparton process is the double parton
collision. Since the momentum scale $p_t$ of a hard interaction
corresponds to much smaller  transverse distances $\sim 1/p_t$ in
coordinate space than the hadronic radius, in a double parton
collision the two interaction regions are well separated in the
transverse space. Also in the c.m. frame pairs of partons from the
colliding hadrons are located in pancakes of thickness $\le (1/x_1
+1 /x_2)/p_{c.m.}$. So two hard collisions occur practically
simultaneously as soon as $x_1, x_2$ are not too small and there
is no cross talk between two hard collisions. A consequence is
that  the different parton processes add incoherently in the cross
section. The double parton scattering cross section, being
proportional to the square of the elementary parton-parton cross
section, is therefore characterized by a scale factor with
dimension of the inverse of a length squared. The dimensional
quantity is provided by the nonperturbative input to the process,
namely by the multiparton distributions. In fact, because of the
localization of the interactions in transverse space, the two
pairs of colliding partons are aligned, in such a way that the
transverse distance between the interacting partons of the target
hadron is practically the same as the transverse distance between
the partons of the projectile. The double parton distribution is
therefore a function of two momentum fractions and of their
transverse distance, and it can be written as $\Gamma(x,x',b)$.
Actually $\Gamma$ depends also on the virtualities of the partons,
$Q^2,Q'^2 $, though to make the expressions more compact we will
not write explicitly this $Q^2$ dependence. Hence the double
parton scattering cross section for the two ``two $\to$ two''
parton processes $\alpha$ and  $\beta$ in an inelastic interaction
between hadrons $a$ and $b$ can be written as:

\begin{eqnarray}
\sigma_D(\alpha,\beta)&=&{m\over2}\int\Gamma_a(x_1,x_2;b)\hat{\sigma}_{\alpha}
x_1,x_1')
\hat{\sigma}_{\beta}(x_2,x_2')\Gamma_b(x_1',x_2';b)dx_1dx_1'dx_2dx_2'd^2b
\label{1}
\end{eqnarray}

\par\noindent
where $m=1$ for indistinguishable parton processes and $m=2$ for
distinguishable parton processes. Note that, though the
factorization approximation of Eq.(\ref{1}) is generally accepted
in the analyses of the multijet processes and appears natural
based on the geometry of the process, no formal proof exists in
the literature. As we will show below the study of the
A-dependence of this process will allow to perform  a stringent
test of this approximation.

In the case of $NN$ scattering one cannot proceed further without
making some simplifying assumptions about transverse correlations
of partons in nucleons. Our key observation is that the
introduction of a new large transverse scale: the  nucleus radius,
allows to separate the effects of the transverse and longitudinal
parton correlations. Essentially, we can express the function
$\Gamma_A(x_1,x_2,b)$ through $\Gamma_N(x_1,x_2,b)$ and the
distribution of nucleons in the nucleus, practically without any
extra model assumption. Here to simplify the discussion we neglect
small non-additive effects in the parton densities, which is a
reasonable approximation for $0.02\leq x \leq 0.5$. In this case
we have to take into account only $b$- space correlations of
partons in individual nucleons. One has therefore two different
contributions to the double parton scattering cross section. The
first one, $\sigma_1^D$, which is represented in Fig,1a, is the
same as for the nucleon target (the only difference being the
enhancement of the parton flux) and the corresponding cross
section is

\begin{equation}
\sigma_1^D=\sigma_D\int d^2BT(B)= A\sigma_D. \label{sigma1}
\end{equation}

\par\noindent
where $T(B)$ is the nuclear thickness, as a function of the impact
parameter of the hadron-nucleus collision $B$.

The contribution to the term in $G_A(x_1',x_2',b)$ due to the
partons originated from different nucleons of the target (Fig.1.b)
can be calculated {\it solely} from the geometry of the problem by
observing that the nuclear density does not change within a
transverse scale $\left<b\right> \ll R_A$.

The two simplest methods are to use the AGK cutting
rules\cite{AGK}, or the technique of \cite{BT}. We can write, for
two indistinguishable parton processes,

\begin{equation}
\sigma_2^D=\frac{1}{2}\int G_N(x_1,x_2)\hat{\sigma}(x_1,x_1')
\hat{\sigma}(x_2,x_2')G_N(x_1')G_N(x_2')dx_1dx_1'dx_2dx_2'\int
d^2BT^2(B),
\label{sigma2}
\end{equation}

\par\noindent
where $G_N(x_1,x_2)=\int d^2b\Gamma_N(x_1,x_2;b),$ while $x_i$ are
nucleon and $x_i'$ are nuclear parton fractions. The distinctive
feature of the $\sigma_2$- term  is  in difference from the case
of $NN$ interactions, since no transverse scale factor related to
the nucleon  scale is present in $\sigma_2^D$. The correct
dimensionality is provided by the nuclear thickness function,
which appears in $\sigma_2^D$ at the second power. The two
contributions $\sigma_1^D$ and $\sigma_2^D$ are therefore
characterized by a different dependence on the atomic mass number
of the target. The $A$-dependence of the two terms is in general a
function of the values of the momentum fractions and of the
virtuality scale of the $2\to 2 $ interactions. The simplest
situation is in the kinematical regime where shadowing corrections
to the nuclear structure function can be neglected. $\sigma_1^D$
is then proportional to $A^1$ and $\sigma_2^D$ to $A^{1.5}$. (Note
that the nuclear surface effects lead to a faster dependence of
$\int T^2(B) d^2B$ on $A$, for $A\leq 240$, than the naive
expectation $A^{4/3}$). The presence of two terms with distinctive
$A$-dependence (and comparable magnitude for a wide range of x,
see Eq.(\ref{adep}) below)  will allow to separate them easely
experimentally and also to check in the course of such an analysis
the factorization approximation of Eq.(\ref{1}).

To estimate the relative importance of $\sigma_1^D$ and
$\sigma_2^D$, and only to that purpose, we use the CDF analysis
\cite{cdf} were it was assumed that all correlations in fractional
momenta may be neglected, so that one can write $\Gamma_N$ in a
factorized form as a product of two parton densities $G_N(x)$ and
of a function of the inter-parton transverse distance $b$:
$\Gamma_N(x,x',b)=G_N(x)G_N(x')F(b)$. (In a sense this could be
considered as marely a convenient parametrization of  the
experimental data). With these simplifications one obtains

\begin{equation}
\sigma_D(\alpha,\beta)=\frac{m}{2}
\frac{\sigma_{\alpha}\sigma_{\beta}}{\sigma_{e ff}},
\label{simplest}
\end{equation}

\par\noindent
where $\sigma_{\alpha}$ and $\sigma_{\beta}$ are the inclusive
cross sections for the two processes $\alpha$ and $\beta$ in the
hadron - hadron interactions. Under the factorization assumption
the whole new information on the hadron structure can be reduced
to a single quantity with dimensions of a cross section,
$\sigma_{eff}$ which was measured by CDF to be:

\begin{equation}
\sigma_{eff}=14.5\pm1.7^{+1.7}_{-2.3}mb.
\label{sigeff}
\end{equation}

Within the accuracy and in the limited kinematic range accessible
to the experiment ($0,01-0.40$ for the photon+jet scattering,
$0.002-0.20$ for the dijet scattering) no evidence was found of a
$x$ dependence of $\sigma_{eff}$, supporting the simplest
uncorrelated picture of the interaction.  However the absolute
value of the cross section is significantly larger (by a factor
$\geq 2$) than the cross section one would obtain within the
factorization hypothesis, when assuming that the transverse
distribution of partons reflects the matter distribution in the
nucleon needed to obtain the value of the nucleon
non-single-diffractive cross section measured by CDF (for
extensive discussions see \cite{DT} ). So the CDF  data  actually
appear to indicate the presence of parton-parton correlations in a
nucleon, though one cannot distinguish whether they  are solely
due to transverse correlations or to a combination of longitudinal
and transverse correlations.

Substituting Eq.(4) in Eq.(2) and the factorized form of
$\Gamma_N$ in Eq.(3) we can estimate the relative importance of
$\sigma_1^D$ and $\sigma_2^D$:

\begin{equation}
{\sigma_2^D\over \sigma_1^D}={\int T^2(B)d^2B \over
A}\sigma_{eff}\approx 0.45\cdot \left({A\over
10}\right)^{0.5}_{\left | A\geq 10\right.}.
\label{adep}
\end{equation}

Here we evaluated $\int T^2(B)d^2B$ by using the standard
experimentally determined Fermi step parameterization of the
nuclear matter densities \cite{Vorobiov}. One can see from
Eq.\ref{adep} that for heavy nuclei, which are available at RHIC,
the second term will constitute about 70\% of the cross section
and hence the study of the A-dependence of the four jet production
will allow a straightforward  separation of the two contributions
to the cross section. (Note also that the cross section of the two
partons $\to$ four jets process, which constitutes a background to
the four$\to$ four processes, depends linearly on A, so that its
contribution may be disentangled by studying the A-dependence of
the cross section). It is worth emphasizing that, if the small
value of $\sigma_{eff}$ is due to the correlation of the
longitudinal distributions, the relative contribution of the
second term would be further enhanced. Considering that the
enhancement of $\sigma_D$ in $p{\bar p}$ collisions is roughly by
a factor two as compared with the naive expectation, one would
expect in this case an additional enhancement of $\sigma_2^D$ by a
factor $\sim \sqrt{2}$.

One could question whether the soft particle production background
may create more serious  problems in $pA$ scattering than in $NN$
scattering. It appears that this problem can be avoided by
choosing $x_i> x_i'$,  $x_i \geq 0.1$ and selecting a kinematics
close to 90 degrees in the c.m. of the partonic collisions. In
this case the jets are produced predominantly in the proton
fragmentation region, where the soft hadron multiplicity in $pA$
collisions is smaller than in $NN$ collisions.

In order to extend the analysis to the  $x_i'\leq 0.01$ kinematics
one needs to take into account shadowing effects in the nuclear
parton densities. Here we restrict our discussion to the case of
the leading twist parton shadowing, which is a pretty safe
approximation for $p_t \geq 5-7 GeV/c$ and $x\geq 10^{-3}$, which
is anyway the minimal cut on $p_t$ to be able to observe the jets.
(For recent estimates of the kinematics were black body/unitarity
effects may become important see \cite{DESY}). In the case where
only one of the nuclear partons is in the shadowing region, the
ratio of $\sigma_2^D/\sigma_1^D$ is modified only by the
dependence of the shadowing on the nuclear impact factor, given
the different selection of impact parameters in two terms
$\sigma_1^D$ and $\sigma_2^D$ (the integration with measure
$T^2(B)$ leads in fact to a somewhat smaller average $B$ as
compared with the integration with measure $T(B)$). We have
performed a numerical estimate of this effect within  the leading
twist approximation of \cite{FS99} and we found that even for
$A=240$ this effect leads to a decrease of $\sigma_2^D/\sigma_1^D$
by $\leq 10\%$. The effect could be in any case studied
experimentally by investigating the single hard scattering as a
function of $A$ in the same kinematics. When both $x_i'$s are in
the shadowing region the evaluation of the effect is more model
dependent, though it still appears to  be rather small. In any
case such kinematics is more appropriate for the study of the
dynamics of the nuclear shadowing and hence is not directly
related to the subject of this letter. Note also  that pushing
such measurements into a kinematical region close to the black
body limit would create additional problems since, due to the
increase of the transverse momenta of the nuclear partons, the
pair-vice azimuthal correlation, which allows an easy
identification of the double parton collision events,  would
become weaker and weaker. Note also that we argued in the
beginning that processes which may violate factorization should be
amplified when $x_i, x_i'$ become smaller. Hence the study of the
A-dependence in this kinematics would provide an additional test
of the factorization approximation.

Summarizing, the study of the $A$-dependence of the double parton
scattering will allow to separate two contributions to the cross
section - due to scattering off one and two nucleons of the
nucleus. Because of the large nuclear size: $R_A\gg R_N$, the
$\sigma_2^D$ term  provides a model independent measurement of the
double parton densities in nucleons - while no such model
independent measurement is possible with proton targets. At the
same time the comparison of the two terms will allow a practically
model independent determination of the transverse separation
between two partons (modulus a possible small effect due to a
different transverse separation of $\Gamma(x_1,x_2,b)$ and
$\Gamma(x_1',x_2',b)$) as well as checking the factorization
approximation.

Obviously one can consider also three parton collisions. In
difference from the case of the double collisions it is more
difficult for the available range on nuclei to extract the triple
parton distribution without making simplifying hypotheses. This is
because the triple scattering process originates due to three
different mechanisms, corresponding to the number of target
nucleons involved. While the terms with one (Fig. 2a) and three
target nucleons (Fig. 2c) are analogous to the contributions
already considered for the double scattering, the contribution
with two different target nucleons - Fig. 2b, is different. In the
latter case, in fact, the integration on the transverse
coordinates of the interacting partons involves at the same time
two partons of the projectile and two partons of the target. The
simplest possibility is that the longitudinal and transverse
degrees of freedom can still be factorized, in this case the
integration over the transverse partonic coordinates gives as a
result a factor with dimensions of the inverse of a cross section.
We call the new dimensional quantity $\sigma_{eff}'$ and a naive
expectation would be that its value is not much different from
$\sigma_{eff}$. The different contributions to the triple
scattering cross section are therefore:

\begin{eqnarray}
\sigma_1^T&=&\sigma_T\int d^2BT(B)=A\sigma_T\nonumber\\
\sigma_2^T&=&\frac{1}{3!}\int G(x_1,x_2,x_3)\hat{\sigma}(x_1,x_1')
\hat{\sigma}(x_2,x_2')\hat{\sigma}(x_3,x_3')dx_1dx_1'dx_2dx_2'dx_3dx_3'
\nonumber\\
&&\times\Bigl[ G(x_1',x_2')G(x_3')+ G(x_2',x_3')G(x_1')+
G(x_1',x_3')G(x_2')\Bigr]\nonumber\\
&&\times\int d^2BT^2(B)\frac{1}{\sigma_{eff}'}\nonumber\\
\sigma_3^T&=&\frac{1}{3!}\int G(x_1,x_2,x_3)\hat{\sigma}(x_1,x_1')G(x_1')G(x_2')G(x_3')\nonumber\\
&&\times\hat{\sigma}(x_2,x_2')\hat{\sigma}(x_3,x_3')dx_1dx_1'dx_2dx_2'dx_3dx_3'
\int d^2BT^3(B),
\label{sigma3}
\end{eqnarray}

\par\noindent
where $\sigma_T$ is the triple parton scattering cross section on
a nucleon target. The second term provides an additional
information about correlations of partons in nucleons while the
third term measures triple parton density in nucleons. If we
assume that the integral over transverse coordinates in dimension
scale for $\sigma_1^T$, $\sigma_2^T$ are approximately
$\sigma_{eff}^{-2}$ and $\sigma_{eff}^{-1}$ we can estimate that
the relative importance of the three terms, for $A\ge 10$, is
approximately:

\begin{equation}
\sigma_1^T:\sigma_2^T:\sigma_3^T=1:1.45(A/10)^{0.5}:0.25(A/10)
\end{equation}

This estimate indicates that the A-dependence of $\sigma^T$ is
much stronger than for $\sigma^D$, with the scattering off several
nucleons becoming important already for light nuclei. The
$\sigma_3^T$ term is likely to become comparable to the other
terms for heavy nuclei, so  in principle an accurate study of the
A-dependence would allow to measure all three terms separately and
hence determine the triple parton density in a nucleon in a model
independent way. Obviously one would need LHC energies and a large
acceptance FELIX-type \cite{FELIX} detector to be able to study
such reactions.

One can go  a step further and try to get information about global
characteristics of the nucleon as a function of the values of the
flavor, $x$'s, etc of the probed partons. Really the number of the
secondaries produced gives indications of the actual transverse
size of the projectile, so that one would expect to observe a
relatively smaller population of sea quarks and gluons in events
with few secondaries, while, in that case, the momentum carried by
the valence should be larger than average. Hence one might start
by studying the correlation between soft characteristics of the
events (the simplest for RHIC would be number of neutrons in the
zero angle calorimeter) and the momentum fraction $x$ of the
projectile parton in a single hard collision \cite{FS85}. A next
step would be to compare the single and double hard scattering
events for fixed $x_1$, say $x_1\sim 0.2$, while increasing the
value of $x_2$. If, when selecting two fast partons in a nucleon,
one selects configurations with a small size, one would expect,
for example, that the number of knockout nucleons would decrease
when $x_2$ increases. Overall, such studies would allow to obtain
unique information about the three-dimensional structure of the
nucleon.

\vskip.25in

{\bf Acknowledgment}

\vskip.15in

We thank L.Frankfurt for useful discussions. We thank the
Department of Energy's Institute for Nuclear Theory at the
University of Washington for hospitality and the Department of
Energy for partial support during the completion of this work.
This work was partially supported by the Italian Ministry of
University and of Scientific and Technological Researches (MURST)
by the Grant COFIN99.

\begin{figure}

\begin{center}

\leavevmode
\epsfxsize=1.0\hsize
\epsfbox{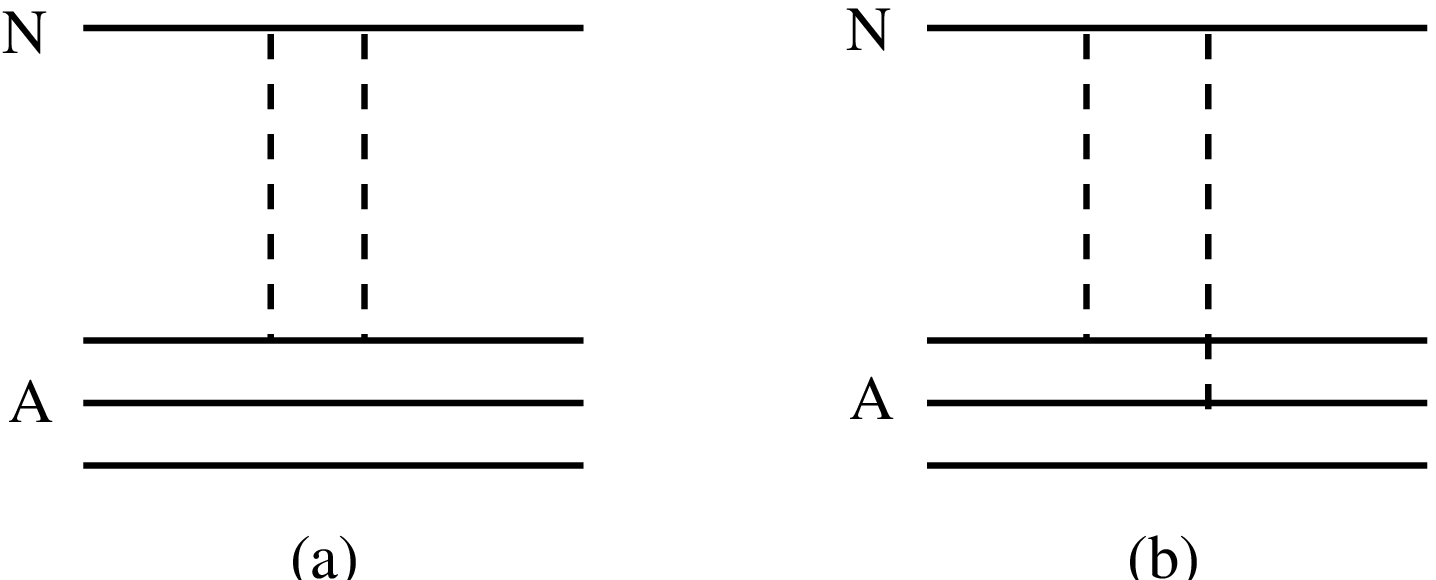}

\end{center}

\caption{Two contributions to the "four $\to$ four" process in
$pA$ scattering. Dashed lines represent hard interactions.}
\label{pa1}
\end{figure}

\newpage

\begin{figure}

\begin{center}

\leavevmode
\epsfxsize=1.0\hsize
\epsfbox{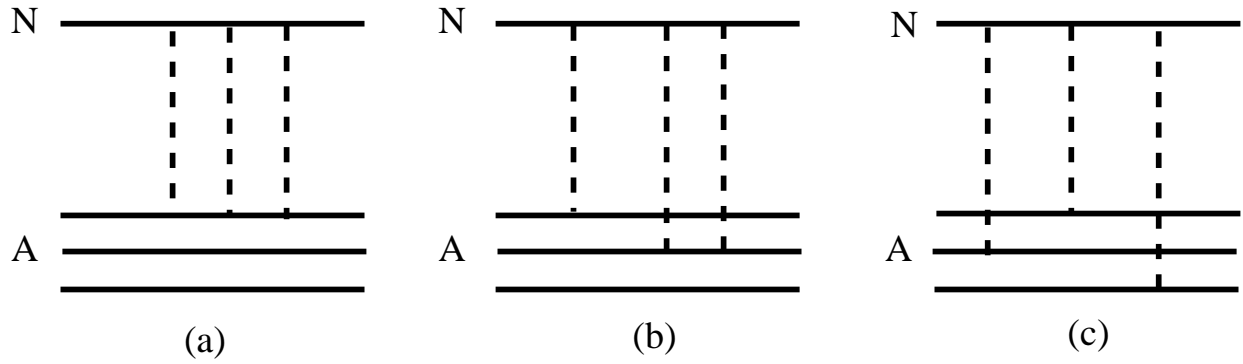}

\end{center}

\caption{Three  contributions to the "six $\to$ six" process in
$pA$ scattering. Dashed lines represent hard interactions.}
\label{pa2}
\end{figure}

\end{document}